\documentstyle[twocolumn,aps,prl,graphicx]{revtex}    

\begin{document}
\bibliographystyle{revtex}
\draft

\title{Novel rapidity dependence of directed flow in high energy heavy ion collisions}

\author{
R.J.M.~Snellings$^{(a)}$, 
H.~Sorge$^{(b)}$, 
S.A.~Voloshin$^{(a)}$\cite{byline},
F.Q.~Wang$^{(a)}$, 
N.~Xu$^{(a)}$}

\address{ (a) Nuclear Science Division, LBNL, Berkeley, California 94720, USA \\
(b) Department of Physics, SUNY Stony Brook, New York 11794, USA}

\date{\today}
\maketitle

\begin{abstract}
For high energy nucleus-nucleus collisions,
we show that a combination of space-momentum correlations
characteristic of radial expansion together with the correlation between
the position of a nucleon in the nucleus and its stopping, results 
in a very specific rapidity dependence of directed flow: 
a reversal of sign in the mid-rapidity region.
We support our argument by RQMD model calculations for Au+Au collisions 
at $\sqrt s$~=~200~$A$GeV.
\end{abstract}


The study of anisotropic flow in high energy nuclear collisions
has attracted increasing attention from both
experimentalists and theorists~\cite{H-G,E877,liu97,posk98}. 
The rich physics of directed and elliptical 
flow~\cite{olli92,lshur,risc96,sorge97,heisel,sorge99,csernai,brachmann,zhang99}
is due to their sensitivity to the system evolution at early time. 
Anisotropic flow in general is also sensitive to the equation of
state~\cite{H-G,sorge99} which governs the evolution of the system 
created in the nuclear collision.

The collective expansion of the system created during a heavy-ion 
collision implies space-momentum correlation in particle distributions
at freeze-out. 
Simplified, this means that particles created on the left 
side of the system move in the left direction and particles created 
on the right side move in the right direction (on average). 
We will show that the rapidity dependence of directed flow
of nucleons and pions can address this space-momentum correlation 
experimentally.

\begin{figure}
\centering
\mbox{
\includegraphics[width=.47\textwidth]{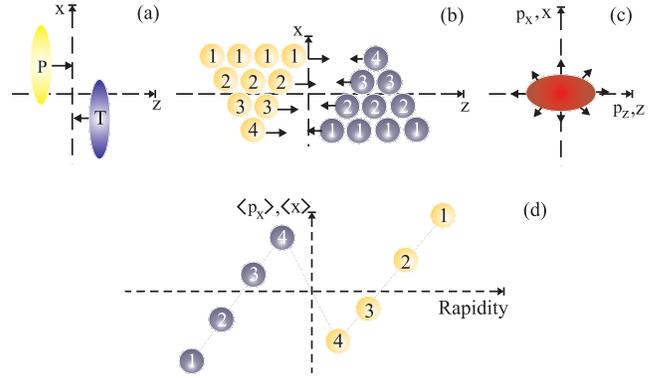}
}
\vspace{.1cm}
\caption{
A schematic sketch of a medium central symmetric heavy-ion collision
in progressing time (a),(c) and the rapidity distribution
of $\langle p_x \rangle$ and $\langle x \rangle$ in (d). 
In (b) the overlap region is magnified and the ``spectators'' are not shown
In these figures $x$ is the
coordinate along the impact parameter direction and $z$ is the coordinate
along the projectile direction (for a more detailed description see text).
}
\label{negativeflow}
\end{figure}

A sketch of a medium central symmetric heavy-ion collision
is shown in Fig.~\ref{negativeflow}, from before the collision (a,b) 
to the resulting distributions of $\langle x \rangle$ and 
$\langle p_x \rangle$ shown in (d).
In Fig.~\ref{negativeflow}a the projectile and target are shown before 
the collision in coordinate space.
In Fig.~\ref{negativeflow}b the overlap region is magnified
and the ``spectators'' are not shown.
It shows in a schematic way the
number of nucleons and their position in the x--z plane
(where $\hat{\bf x}$ is the impact parameter direction).
Projectile nucleons (light color) at negative $x$ suffer more rapidity 
loss than those at positive $x$, ending up closer to mid-rapidity.
Assuming a positive space-momentum correlation (as indicated in 
Fig.~\ref{negativeflow}c), these nucleons have negative $\langle p_x \rangle$, 
while those at positive 
$x$ have positive $\langle p_x \rangle$. 
This results in a wiggle structure in the rapidity dependencies of
$\langle x \rangle$ and $\langle p_x \rangle$ which is 
shown schematically in Fig.~\ref{negativeflow}d.

The shape of the wiggle, both the magnitude of
$\langle p_x \rangle$ and the rapidity range, depends on
the strength of the space-momentum correlation, the initial beam-target 
rapidity gap and the amount of stopping.
Therefore, the dependence of the wiggle
on the collision centrality, system size and center of mass energy
may reveal important information on the 
relation between radial flow and baryon stopping.
In addition, it has been shown that the magnitude of $\langle p_x \rangle$
depends on the nuclear matter equation of state~\cite{sorge99a}.

The above picture changes for collisions at lower energies,
where there is no clear rapidity separation between projectile and target
nucleons at freeze-out, because nucleons cross over mid-rapidity. 
Moreover, when the time for the nuclei to pass each other becomes long
relative to the characteristic time scale for particle production, the
interactions between particles and spectators (so-called shadowing) 
become important. This has been pointed out by many people and
recently in \cite{sorge97,bass,liu99}. 
This is consistent with the results of heavy ion collisions, in the 2 to 
158~$A$GeV energy range, where the experimental observed slope around
mid-rapidity in directed flow shows a trend from a positive value at 
2~$A$GeV~\cite{liu97} to almost zero at 158~$A$GeV~\cite{posk98}.

Note that the change of sign of directed flow at mid-rapidity
has been discussed for Ca+Ca collisions at 350~$A$MeV~\cite{soff}
and for Au+Au collisions at 11~$A$GeV~\cite{csernai,brachmann,bravina}.
However, the physical origins on which these predictions are based
are different from what we discuss in this Letter. 
In 350~$A$MeV Ca+Ca collisions the wiggle originates
from a combination of repulsive nucleon-nucleon collisions
and an attractive mean field.
The repulsive nucleon-nucleon collisions dominate at the 
mid-rapidity region and lead to a positive slope for 
$\left\langle p_x \right\rangle$ versus rapidity.
The attractive mean field dominates at beam and target rapidities
and leads to a negative slope. 
In 11~$A$GeV Au+Au collisions the wiggle is caused by the
longitudinal hydrodynamic expansion of a tilted source~\cite{csernai}.
It has been noticed that in hydro calculations the wiggle only appeared 
if a QGP equation of state is used. 
The QGP equation of state is a prerequisite to reach the stopping
needed to create this tilted source. 
The predicted wiggle in this Letter does not assume a
QGP equation of state. The other main difference is that in our
prediction we specifically assume in-complete stopping. 

The arguments used in this Letter which lead to a
change of sign of directed flow at mid-rapidity are valid 
on general grounds. However, to test the picture quantitatively we
study Au+Au collisions at $\sqrt s$~=~200~$A$GeV in an impact 
parameter range $b$=5--10~fm, using the Relativistic Quantum Molecular
Dynamics (RQMD v2.4) model in cascade mode~\cite{rqmd}. 

\begin{figure}
\centering\mbox{
\includegraphics[width=.48\textwidth]{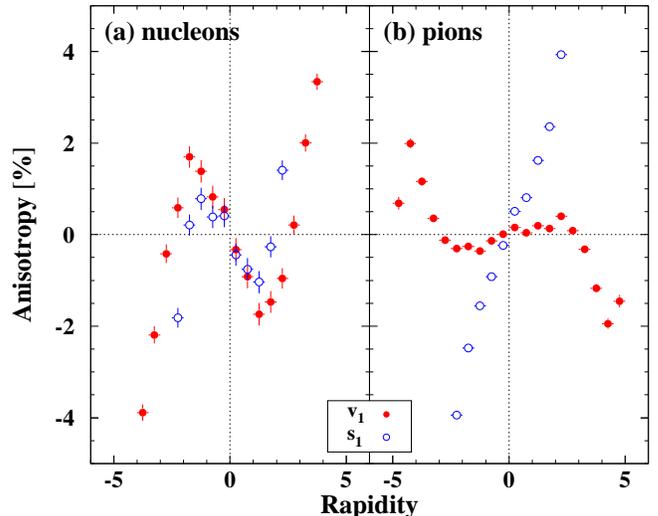}}
\caption{RQMD calculations of $v_1$ (filled circles) and $s_1$ (open
circles) for nucleons (left panel) and pions (right panel). }
\label{pr}
\end{figure}

To characterize directed flow, we use the first Fourier
coefficient~\cite{posk98,volosz}, $v_1$, of the 
particle azimuthal distribution.
At a given rapidity and transverse
momentum the coefficient is determined by $v_1 =
\left\langle \cos \phi \right\rangle$, where $\phi$ is the azimuthal
angle of a particle relative to the reaction plane angle (${\bf \hat{x}}$ 
direction in Fig.~\ref{negativeflow}). 
Similarly a Fourier coefficient can be determined in coordinate space, 
$s_1 = \left\langle \cos \phi_s \right\rangle$ ~\cite{liu99},
where $\phi_s$ is the azimuthal angle of a particle, determined from 
the freeze-out coordinates $x$ and $y$, relative to the reaction plane angle. 
Figure~\ref{pr} shows the
RQMD calculations of $v_1$ and $s_1$ for nucleons and pions in Au+Au
collisions at RHIC energy. Indeed, the shape at mid-rapidity for
nucleons is consistent with the picture described above: both $v_1$
and $s_1$ show a negative slope at mid-rapidity. For larger rapidities
the $s_1$ values leave the ordinate scale.

Pions are produced particles and their space-rapidity correlation is different
from that of nucleons shown in Fig.~~\ref{negativeflow}d.
Due to the asymmetry in the numbers of colliding target and
projectile nucleons, the pions produced at positive $x$ 
shift toward positive rapidity.
The pions produced at negative $x$ shift toward negative rapidity. 
This results in a positive space-rapidity correlation without a wiggle, 
see Fig.~\ref{pr}b. (open circles).
Due to the space-momentum correlation, the momentum 
distribution tends to follow the space distribution. This leads to the
positive slope of $v_1$ at mid-rapidity for the pions, see Fig.~\ref{pr}b.
(filled circles).
However, shadowing by nucleons is also important in the formation of pion 
directed flow.
The contribution is relatively small in the mid-rapidity region, where
in high energy nucleus-nucleus collisions the nucleon to pion ratio
is small. At beam/target rapidity the shadowing becomes dominant, this
explains why $s_1$ has the opposite sign from $v_1$ for pions close
to beam/target rapidity.

In this Letter we have shown that the combination of space-momentum 
correlations characteristic of radial expansion together with the 
correlation between the position of a nucleon in the nucleus and 
its stopping, results in a wiggle in the rapidity dependence of 
directed flow in high energy nucleus-nucleus collisions. 
Moreover, the amount of stopping and the space-momentum 
correlation depend on the equation of state and this affects the 
strength of the wiggle around mid-rapidity~\cite{sorge99a}.
Finally, because the wiggle appears at mid-rapidity,
it is accessible by the current SPS experiment NA49 and 
the near future RHIC experiments. The study of its dependence on
collision centrality, system size and the center of mass energy
may reveal important information on the relation between collective 
radial flow and baryon stopping.

We are grateful to G.E. Cooper, Y. Pang, S. Panitkin, A.M. Poskanzer, 
G. Rai, H.G. Ritter and H.~Str\"obele for useful discussions. 
This work was supported by the Director, Office of Energy Research, 
Office of High Energy and Nuclear Physics, 
Division of Nuclear Physics of the U.S. Department
of Energy under Contract DE-AC03-76SF00098.

\end{document}